\documentclass[11pt,a4paper]{article}
\usepackage{amsmath,latexsym,fullpage,amssymb,color}
\usepackage{epic,eepic,ifthen,graphics,epsfig}
\usepackage[english]{babel}
\usepackage{times}
\usepackage{float}
\usepackage{xspace}
\usepackage[T1]{fontenc}
\newlength {\squarewidth}

\newtheorem{theorem}{Theorem}

\newtheorem{property}{Property}

\newcommand{\toto}{xxx}
\newenvironment{proofT}{\noindent{\bf Proof }}
{\hspace*{\fill}$\Box_{Theorem~\ref{\toto}}$\par\vspace{3mm}}

\newenvironment{lemma-repeat}[1]{\begin{trivlist}
\item[\hspace{\labelsep}{\bf\noindent Lemma~\ref{#1} }]}%
{\end{trivlist}}

\newenvironment{theorem-repeat}[1]{\begin{trivlist}
\item[\hspace{\labelsep}{\bf\noindent Theorem~\ref{#1} }]}%
{\end{trivlist}}

\newcounter{linecounter}
\newcommand{\linenumbering}{\ifthenelse{\value{linecounter}<10}
{(\arabic{linecounter})}{(\arabic{linecounter})}}
\renewcommand{\line}[1]{\refstepcounter{linecounter}\label{#1}\linenumbering}
\newcommand{\resetline}[1]{\setcounter{linecounter}{0}#1}
\renewcommand{\thelinecounter}{\ifnum \value{linecounter} > 
9 \else \fi\arabic{linecounter}}
\newfloat{algorithm}{thp}{lop}
\floatname{algorithm}{Algorithm}
\newfloat{construction}{thp}{lop}
\floatname{construction}{Construction}
\newcommand{\Xomit}[1]{}
\newcommand{\KREG}{\mathit{KREG}}
\newcommand{\propose}{{\sf propose}}
\newcommand{\RWk}{RW$_{k}$}  

\newcommand{\RWkprime}{RW$_{k'}$}
\newcommand{\RWone}{RW$_{1}$}
\newcommand{\wwrite}{{\sf write}}  
\newcommand{\rread}{{\sf read}}  
\newcommand{\return}{{\sf return}} 

\begin{document}

\title{\bf  A Simple Object that Spans the Whole Consensus Hierarchy}

\author{Achour Most\'efaoui$^{\dag}$, Matthieu Perrin$^{\circ}$, 
        Michel Raynal$^{\star,\ddag}$\\~\\
$^{\dag}$LINA, Universit\'e de Nantes, 44322 Nantes, France \\
        $^{\circ}$Computer Science Department, The Technion, Haifa, Israel \\
$^{\star}$Institut Universitaire de France\\
$^{\ddag}$IRISA, Universit\'e de Rennes, 35042 Rennes, France \\
}

\date{}

\maketitle


\begin{abstract}
This paper presents a simple generalization of the basic atomic
read/write register object, whose genericity parameter spans the whole
set of integers and is such that its $k$-parameterized instance has
exactly consensus number $k$. This object, whose definition is pretty
natural, is a sliding window register of size $k$.  Its interest lies in
its simplicity and its genericity dimension which provides a global
view capturing the whole consensus hierarchy.
Hence, this short article must be seen as a simple pedagogical
introduction to Herlihy's consensus hierarchy.

%
~
\end{abstract}

\section{Wait-free Computing Model and the Consensus Hierarchy}

\paragraph{Crash-prone asynchronous read/write-based systems}
This paper considers the classical distributed computing model called 
{\it read/write wait-free}  model~\cite{H91}.
It is  composed of a set of $n$ sequential processes denoted $p_1$, ..., $p_n$,
which communicate through atomic read/write registers~\cite{HRR13,HW90,L86,M86}.

Each process is asynchronous,  which means
that it proceeds at its own speed, which can be arbitrary and remains
always unknown to the other processes, and   
executes its local algorithm until it  possibly crashes, where a  
crash is a premature halt.  Any number of processes may crash in a run, 
and after crashing a process does not recover. 
A process that crashes in a run is said to be {\it faulty}. Otherwise, 
it is {\it correct} or {\it non-faulty}.
Let us notice that, due to process crashes and asynchrony, no process can
know if an other process crashed or is only very slow.

\paragraph{Consensus object}
The notion of  a {\it universal} object with respect to fault-tolerance 
was  introduced by M. Herlihy~\cite{H91}. 
An object type $T$ is {\it universal} if it is possible to implement any 
object (defined by a sequential specification) in the read/write wait-free
model enriched with any number of objects of type $T$.  An algorithm providing 
such an implementation is called a {\it universal construction}. 
It is shown in~\cite{H91} that {\it consensus} objects are universal.  
These objects allow the processes to propose values and agree on one of them. 
More precisely, such an object provides the processes with a single operation, 
denoted $\propose()$, that a process can invoke only once, and returns it a 
value.  When $p_i$ invokes  $\propose(v_i)$,we say that it ``proposes the 
value $v_i$'', and if $v$ is the returned value we say that it ``decides $v$''. 
The consensus object is defined by the three following properties:
\begin{itemize}
\vspace{-0.2cm}
\item Validity. 
If a process decides a value, this value was proposed by a process. 
\vspace{-0.2cm}
\item Agreement. 
No two processes decide different values. 
\vspace{-0.2cm}
\item Termination. 
If a correct process invokes $\propose()$, it decides a value. 
\end{itemize}

Termination states that if a correct process invokes $\propose()$, it
decides a value whatever the behavior of the other processes
(wait-freedom progress condition).  Validity connects the output to the 
inputs, while Agreement states that the processes cannot decide
differently. A sequence of consensus objects is used in the following
way in a universal construction.  According to its current view of the
operations invoked on, and not yet applied to, the object $O$ of type $T$ 
that is built, each process proposes to the next consensus instance
a sequence of operations to be applied to $O$, and the winning
sequence is actually applied.  An helping mechanism~\cite{CPT15} is
used to ensure that all the operations on $O$ by any correct process are
eventually applied to $O$.

\paragraph{Consensus numbers and consensus hierarchy}
The notion of a {\it consensus number} associated with an object type
$T$ (denoted CN($T$) in the following) was introduced by Herlihy
in~\cite{H91}.  It is the greatest integer $n$ such that consensus can
be implemented in a system of $n$ processes with atomic read/write
registers and objects of type $T$.  If there is no such finite $n$,
the consensus number of $T$ is $+\infty$.  Hence, a type $T$ such that
CN$(T)\geq n$ is universal in a system of $n$ (or less) processes.

It appears that the consensus numbers define an infinite hierarchy
(Herlihy's hierarchy) in which atomic read/write registers
have consensus number $1$, object types such as Test\&Set,
Fetch\&Add, and Swap, have consensus number 2, etc., until object
types such as Compare\&Swap, Linked Load/Store Conditional (and a few
others) that have consensus number $+\infty$.  In between, read/write
registers provided with $m$-assignment\footnote{Such an
assignment updates atomically $m$ read/write registers. It is sometimes
written $X_1,X_2,\cdots,X_m \leftarrow v_1,\cdots,v_m$ where the
$X_i$ are the registers, and each $v_i$ the value assigned to $X_i$.}
with $m>1$, have consensus number $(2m-2)$.
(Recent developments on synchronization objects and 
consensus numbers can be found in~\cite{AEG16,CPT15,IR10}.)

\paragraph{Content of the paper}
This paper addresses the following question: Does it exist a simple
object family, parameterized by an integer $k$, that covers the whole
consensus hierarchy (i.e., whose object instantiated with number $k$
has exactly consensus number $k$)?  The paper answers positively this
question by presenting a simple object family, 
and shows that, for any $k\geq 1$, its $k$-parameterized instance has consensus 
number $k$. This object is a very simple and natural generalization of 
the most basic  computing object, namely the atomic read/write register,
extended to become a sliding window register of size $k$.  This object
family has two noteworthy properties. One is its simplicity. The other
one lies in the fact that (to our knowledge) it is the only generic 
object spanning {\it all} consensus numbers. This has several advantages,
among which, its pedagogical dimension (easy to understand and teach
to students), its universality dimension (no need to introduce a
specific object at each level of the consensus hierarchy to capture
it), and its definition itself (a simple and natural generalization of
an atomic read/write register).

\section{The Atomic $k$-Sliding  Read/Write Register (\RWk)}

\paragraph{Definition}
As previously indicated, a $k$-{\it sliding  read/write register}
(in short  \RWk) is a natural generalization of an atomic read/write register,
which corresponds to the case $k=1$.  
Let $\KREG$ be such an object. It can be seen as a sequence of values, 
accessed by two atomic operations denoted $\KREG.\wwrite()$ and 
$\KREG.\rread()$. ``Atomic'' means that these operations appear as if 
they have been executed in some sequential order, and this total order is 
such that, if operation ${\sf op1}$ terminates before operation ${\sf op2}$ 
starts, then  ${\sf op1}$ appears before  ${\sf op2}$ \cite{HW90,L86,M86}.

The invocation of $\KREG.\wwrite(v)$ by a process adds the value $v$
at the end of the sequence $\KREG$, while an invocation of
$\KREG.\rread()$ returns the ordered sequence of the last $k$ written
values (if only $x<k$ values have been written, the default value
$\bot$ replaces each of the $(k-x)$ missing values).

Hence,  an {\RWk} object is a sequence containing all the values that have 
been written (in their atomicity-defined writing order),  
and whose each read operation returns the $k$ values that 
have been  written just before it, according to the atomicity order. 
As already indicated, it is easy to see that, for $k=1$,  {\RWk} is a 
classical atomic read/write register. For $k=+\infty$, 
each read operation returns the whole sequence of values written so far.  
Let us notice that {\RWk}  objects appear in some applications 
(e.g., the object that models the content of a screen in an instant messenging 
service where only the last $k$ received messages are displayed, or the screen 
describing plane time departures in airports~\cite{P16}).\footnote{An object 
  close to  {\RWk}  objects was concurrently and independently introduced
  in~\cite{EGSZ16} to address complexity issues in the 
  context of multiprocessor synchronization.}

\paragraph{Ranking the objects of the $\{$\RWk$\}_{k\geq 1}$ family}
Let \RWk~$\geq$  \RWkprime~denotes the fact that 
an \RWkprime~object can be built from an  \RWk~object.
The following property follows directly the length of the sequences
returned by these objects. 

\begin{property}
\label{ranking-property}
$\forall k, k': (k\geq k') \Rightarrow  
({\mbox{\em\RWk}} \geq {\mbox{\em\RWkprime}})$.
\end{property}

\section{The Consensus Number of \RWk ~is $\geq k$}

This section shows that the consensus number of  an \RWk~object is at least
$k$. To this end, Algorithm~\ref{algo:consensus} 
builds a consensus object for $k$ processes from an \RWk~object $\KREG$.

\begin{algorithm}[h!]
\centering{\fbox{
\begin{minipage}[t]{150mm}
\footnotesize 
\renewcommand{\baselinestretch}{2.5}
\resetline
\begin{tabbing}
aaaaa\=aaa\=aaaaa\=aaaaaa\=\kill

{\bf operation} $\propose(v_i)$ {\bf is}\\

\line{algo-cons-01} 
\>  $\KREG.\wwrite(v_i)$\\

\line{algo-cons-02} 
\>  $seq_i \leftarrow \KREG.\rread()$;\\

\line{algo-cons-03}
\> {\bf let} $d$ {\bf be} the first non-$\bot$ value in $seq_i$;\\

\line{algo-cons-04}
\>  $\return(d)$\\

{\bf end operation}.
\end{tabbing}
\end{minipage}
}
\caption{Solving consensus from an \RWk~object (code for $p_i$)}
\label{algo:consensus}
}
\end{algorithm}

\begin{theorem}
\label{theo:geq-k}
{\em CN(\RWk)} $\geq k$.
\end{theorem}

\begin{proofT}
Let us consider a read/write wait-free system of $k$ processes.
The consensus Termination property follows from the Termination
properties of the operations $\KREG.\wwrite()$ and $\KREG.\rread()$ of
the underlying atomic object $\KREG$ (lines~\ref{algo-cons-01}
and~\ref{algo-cons-02}), and the fact that the algorithm contains
neither loops, nor  wait statements.

As at most $k$ processes invoke the consensus operation $\propose()$,
the underlying object $\KREG$ contains at most $k$ values.  Moreover,
the oldest of them is the value $v$ written by the first process that
executed $\KREG.\wwrite()$ (line~\ref{algo-cons-01}).  It follows that
the value extracted (line~\ref{algo-cons-03}) from its local sequence
$seq_i$ by any process $p_i$ is $v$, which proves the consensus
Agreement property.  The proof of the consensus Validity property
follows from the same reasoning.  
\renewcommand{\toto}{theo:geq-k}
\end{proofT}

\section{The Consensus Number of \RWk ~is $< k+1$}

This section shows that, for any finite value $k$, the consensus number 
of  an \RWk~object is  smaller than $(k+1)$.  
The proof is a simple adaptation of impossibility proofs 
found in textbooks (such as~\cite{AW04,L96,R13,T06}), which all rest
on the basic  concepts (e.g., notion of valence) and techniques 
introduced in~\cite{FLP85} in the context of message-passing systems 
(and then used in~\cite{LA87} in the context of wait-free read/write systems).

\paragraph{Definitions} (The definitions that follow are from~\cite{FLP85}.)
Without loss of generality, the proof considers binary consensus, i.e., 
only the values $0$ and $1$ can be proposed by the processes
(there are  algorithms that implement multivalued consensus on top 
of binary consensus~\cite{R13}). 

A configuration is a global state made up of the local states of each 
process and the state of every object shared by the processes. 
In our case, as \RWk $\geq$ \RWone (Property~\ref{ranking-property}), 
we consider that the only objects shared by the processes are 
\RWk~objects. 

Assuming an algorithm  $A$ implementing a consensus object, 
a configuration $\Sigma$ attained by an execution of $A$ is $v$-valent 
($v\in\{0,1\}$), if only the value $v$ can be decided from $\Sigma$. 
Such configurations are said to be {\it monovalent}. Otherwise, 
they are said to be {\it bivalent} (the dices are not yet cast!). 
Let us observe that there is an initial configuration that is 
bivalent\footnote{Assume $p_i$ proposes $0$, while $p_j$ proposes $1$. 
It follows from the consensus Validity property that, 
if all the processes except $p_i$ crash initially, only $0$ can be decided.
Similarly,  if all the processes except $p_j$ crash initially, only $1$ 
can be decided.  It follows that the corresponding initial configuration
is bivalent.}. Moreover, let us notice that -due to its very definition- 
any configuration that follows a $v$-valent configuration is $v$-valent. 

A schedule $\sigma$ is a sequence of operations on shared objects
issued by the processes. Let us observe that, given an initial
configuration, any consensus algorithm $A$ must terminate (all correct
processes must decide).  Consequently all the schedules it can produce
(whatever the failure and asynchrony pattern) must eventually attain 
a monovalent configuration.

$\Sigma$ being a configuration, let 
${\sf op}_x(\Sigma)$ denotes the configuration attained from $\Sigma$ 
by executing ${\sf op}_x$ (the next read or write operation 
on a \RWk~object issued by $p_x$), and $\sigma(\Sigma)$ be the configuration 
attained from  $\Sigma$ by executing the schedule~$\sigma$. 

A {\it maximal} bivalent schedule is a schedule that ends in a bivalent
configuration $\Sigma$ such that the next operation issued by any
process produces a monovalent configuration. Let us notice that, if
there is an algorithm solving consensus, any of its executions has a
maximal schedule (otherwise $A$ will have non-terminating executions).

\begin{theorem}
\label{theo:leq-k}
Let $k <\infty$.
{\em CN(\RWk)} $ < k+1$.
\end{theorem}

\begin{figure}[th]
\centering{
\hspace{-1.5cm}
\scalebox{0.37}{\input{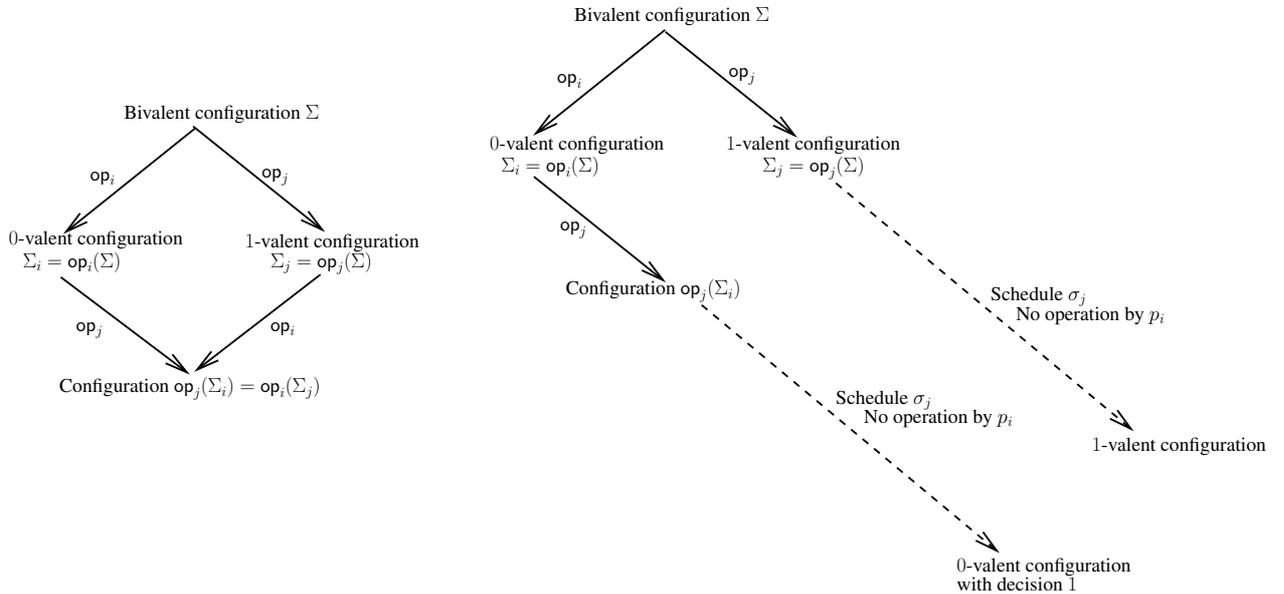}}
\caption{Schedule illustrations}
\label{fig:proof-cases}
}
\end{figure}

The proof can be seen as a straightforward generalization of the proof given 
in~\cite{LA87}, which  shows that atomic registers (i.e., \RWone~registers)
have consensus number $1$.

\begin{proofT}
As in~\cite{FLP85}, starting with an algorithm $A$ assumed to
implement consensus, and an initial bivalent configuration, the proof
consists in building an execution of $A$ in which there is no maximal
schedule.  Consequently, all its configurations are bivalent, from
which follows that the schedule is infinite: $A$ does not satisfy
the consensus Termination property.

Hence, let us consider a read/write wait-free system of $(k+1)$
processes, enriched with any number of \RWk~objects. As $A$ is assumed to
terminate, each of its executions generates a maximal schedule, i.e.,
produces a bivalent configuration $\Sigma$ after which there is no
more bivalent configurations.  The proof is a classical case analysis
depending on whether the next operation issued by each process is a
read or write operation, and whether they are on the same or
different \RWk~objects.  Let $p_i$ and $p_j$ be two processes whose
next operations to execute in $\Sigma$ are ${\sf op}_i$ and  ${\sf op}_j$, 
producing the $0$-valent configuration $\Sigma_i= {\sf op}_i(\Sigma)$, and 
the $1$-valent configuration $\Sigma_j= {\sf op}_j(\Sigma)$, respectively.

\begin{itemize}
\vspace{-0.2cm}
\item  Case 1 (same as Lemma 1 in~\cite{FLP85}, left size of 
Figure~\ref{fig:proof-cases}): 
The operations ${\sf op}_i$ and  ${\sf op}_j$ are on different \RWk~objects. 
We have then ${\sf op}_j({\sf op}_i(\Sigma))={\sf op}_i({\sf op}_j(\Sigma))$
(being on different objects, the operations commute without side effect), 
from which we conclude that this configuration is bivalent, which contradicts  
the fact that $\Sigma$ is maximal.
\vspace{-0.2cm}
\item Case 2:  
The next operations ${\sf op}_i$ and ${\sf op}_j$ issued by $p_i$ and $p_j$
are on the same \RWk~object and one of them (e.g., ${\sf op}_i$) is a read. 
In this case,  there is a schedule $\sigma_j$, starting from the 
$1$-valent configuration
$\Sigma_j={\sf op}_j(\Sigma)$, in which all the processes except $p_i$
(which stops for an arbitrary long period or crashes)  issue operations 
and eventually decide. As $\Sigma_j={\sf op}_j(\Sigma)$ is $1$-valent, 
they decide $1$. 

Let us now consider  ${\sf op}_j(\Sigma_i)={\sf op}_j({\sf op}_i(\Sigma))$. 
This configuration differs from  
$\Sigma_j={\sf op}_j(\Sigma)$ only in the local 
state of $p_i$ (which read the \RWk~object in the configuration
$ {\sf op}_j(\Sigma_i)= {\sf op}_j({\sf op}_i(\Sigma))$,
while it does not in $\Sigma_j={\sf op}_j(\Sigma)$)
See an illustration on the  right size of Figure~\ref{fig:proof-cases}.
Let us  apply the schedule $\sigma_j$ 
to  configuration ${\sf op}_j(\Sigma_i)={\sf op}_j({\sf op}_i(\Sigma))$. 
This is possible because no process (except $p_i$) can distinguish 
${\sf op}_j({\sf op}_i(\Sigma))$ from  ${\sf op}_j(\Sigma)$.
From the schedule  $\sigma_j$, it follows that $p_j$ decides $1$, 
contradicting the fact that the  configuration 
$\Sigma_i={\sf op}_i(\Sigma)$ is $0$-valent. 
\vspace{-0.2cm}
\item 
 Case 3: 
In $\Sigma$, the next operation by each process is a write, 
and these write operations are on the same
\RWk~object $\KREG$ (\footnote{The intuition that underlies
this case is the following.  While $p_i$ can be the first process
that writes a value (say $0$) in $\KREG$ (thereby producing a
$0$-valent configuration) and then pauses for an arbitrarily long
period, it is possible that the next process writes $1$, and the
$(k-1)$ other processes write also a value, whose net effect is the
elimination of the value written by $p_i$ from the current window.}).
The reasoning is similar to  Case 2. 
Let $\Sigma_i={\sf op}_i(\Sigma)$ be 
$0$-valent, and   $\Sigma_j={\sf op}_j(\Sigma)$ be $1$-valent.
Let $\sigma_j$ be a schedule, starting  from $\Sigma_j$ in which 
\begin{itemize}
\vspace{-0.2cm}
\item  (a)  the first  $(k-1)$ operations are the write of  $\KREG$
invoked by the $(k-1)$ processes different from $p_i$ and $p_j$.
\vspace{-0.1cm}
\item (b) all processes, except $p_i$, 
execute steps until each of them  decides, and
\vspace{-0.1cm}
\item (b) $p_i$ executes no operation.
\end{itemize}
Let us notice that such a schedule is possible because, in $\Sigma$, the
next operation of each process is a write into  $\KREG$ 
(Case assumption, which implies item (a)\footnote{The important point  is
here  the following: in $\sigma_j$ no process different from $p_i$
can know the value written in $\KREG$ by $p_i$.}), and 
the algorithm $A$ terminates (hence each correct process 
invokes the consensus operation and decides, which implies item (b)).

Let ${\sf op}_j\sigma_j$ denote the schedule composed 
of ${\sf op}_j$ followed by $\sigma_j$. 
As  $\Sigma_j ={\sf op}_j(\Sigma)$ is $1$-valent, all processes 
involved in ${\sf op}_j\sigma_j$ (i.e., all processes except $p_i$) decide $1$. 

Let us now consider the monovalent state  $\Sigma_i$, 
in which $p_j$ applies ${\sf op}_j$. Let us observe that no
process, except $p_i$, can distinguish $\Sigma_j$ from ${\sf op}_j(\Sigma_i)$
(they have the same local states in both). 
It follows that the schedule  ${\sf op}_j\sigma_j$  
(executed previously from  $\Sigma$)   can also be executed from $\Sigma_i$. 
The first $k$ operations of this  schedule are 
a write operation on $\KREG$ issued by each process different 
from $p_i$. Moreover, at the end of this schedule, 
all the processes (except $p_i$, which is not involved in ${\sf op}_j\sigma_j$) 
decide $1$. This contradicts the fact that $\Sigma_i$ is $0$-valent, 
which concludes  the proof. 
\end{itemize}
\vspace{-0.4cm}
\renewcommand{\toto}{theo:leq-k}
\end{proofT}
\section{Conclusion}
\label{sec:conclusion}

This paper first introduced a new type of concurrent object,
parameterized by an integer $k$, namely an atomic read/write sequence
which can be accessed by a read and a write operation.  Each write
adds a new value at the end of the sequence, while a read returns the
last $k$ written values. This generic object, called $k$-sliding
 read/write register, has an instance for each integer $k$.  The
instance $k=1$ corresponds to the classical atomic read/write
register, which is the most basic object of computing.  Then, the
paper has shown that the consensus number of such a $k$-parameterized
object is $k$.  Hence, this object family covers the whole spectrum of
Herlihy's consensus hierarchy, a noteworthy pedagogical property. 
From a technical point of view, this result may help better understand 
the synchronization power of concurrent objects. Moreover, it is sufficient to
show that an object  can be implemented with a $k$-sliding
read/write register to prove its  consensus number is at most $k$.

\vspace{-0.15cm}
\section*{Acknowledgments}
\vspace{-0.15cm}
This work has been partially supported by the  French  ANR  project 
DESCARTES 
devoted to layered and modular structures in distributed computing. 


\end{document}